\documentclass{article}
\usepackage{amsfonts}

\usepackage{amsmath}

\begin{document}

\title{Entanglement-assisted capacity of constrained channels\thanks{
Work partially supported by INTAS grant 00-738.}}
\author{A. S. Holevo \\
Steklov Mathematical Institute, 119991 Moscow, Russia}
\date{}
\maketitle

\begin{abstract}
In this paper we fill a gap in previous works by proving the conjectured
formula for the classical entanglement-assisted capacity of quantum channel
with additive constraint (such as Bosonic Gaussian channel). The main tools
are the coding theorem for classical-quantum constrained channels and a
finite dimensional approximation of the input density operators for
entanglement-assisted capacity. We also give  sufficient conditions under
which suprema in the capacity formulas are achieved.
\end{abstract}

\bigskip The formula for the entanglement-assisted capacity of a noisy
quantum channel, expressing it as the maximum of mutual information over
input states, was obtained in \cite{eac}, \cite{eac2} for channels in finite
dimensional Hilbert space. Alternative proof was given in \cite{heac}. In
\cite{eac2}, \cite{hw} the appropriately modified formula was also applied
to quantum Gaussian channels which are the most important example of
constrained channels in infinite dimensions. In this paper we fill the gap
in previous works by proving the conjectured formula (\ref{eaco}). The main
tools are the coding theorem for classical-quantum constrained channels and
a finite dimensional approximation of the input density operators for
entanglement-assisted capacity. We also give sufficient conditions under
which suprema in the capacity formulas are achieved.

\bigskip \textbf{1.} We first consider the case of classical-quantum (c-q)
channel with infinite alphabet $\mathcal{X}=\{x\}$. For every $x$ let $S_{x}$
be a density operator in a Hilbert space $\mathcal{H}$ (in general, infinite
dimensional) with finite von Neumann entropy $H(S_{x})$. The c-q channel is
given by the mapping $x\rightarrow S_{x}.$

Let $f(x)$ be a nonnegative nonconstant function defined on the input
alphabet. Passing to block coding, we put the additive constraint onto the
input words $w=(x_{1},...,x_{n})$ by asking
\begin{equation}
f(x_{1})+\ldots +f(x_{n})\leq nE,  \label{1a-2}
\end{equation}
where $E$ is a positive constant. The classical capacity of such channels
was defined and computed in \cite{hol} under a condition of uniform
boundedness of the entropies $H(S_x)$. This condition is not suitable for
our purpose here, and by using almost the same argument we can prove

\textbf{Proposition 1}. Denote by $\mathcal{P}$ the class of finite input
distributions $\pi =\{\pi _{x}\}$ on $\mathcal{X}$ satisfying
\begin{equation}
\sum_{x}\pi _{x}f(x)\leq E.  \label{1a-1}
\end{equation}
We assume $\mathcal{P}$ is nonempty and impose the following condition onto
the channel:
\begin{equation}
\sup_{\pi \in \mathcal{P}}H\left( \sum_{x}\pi _{x}S_{x}\right) <\infty .
\label{1-62}
\end{equation}
The classical capacity of the channel $x\rightarrow S_{x}$ under the
constraint (\ref{1a-2}) is finite and given by

\begin{equation}
C=\sup_{\pi \in \mathcal{P}}\left[ H\left( \sum_{x}\pi _{x}S_{x}\right)
-\sum_{x}\pi _{x}H\left( S_{x}\right) \right] .  \label{1a-9}
\end{equation}

Let the input alphabet $\mathcal{X}$ be a locally compact subset of a
separable metric space (e. g. a closed finite dimensional manifold, or a
discrete countable set, in which case the integrals below should be
understood as sums). Consider the channel given by \textit{weakly continuous}
mapping $x\rightarrow S_{x}$ from the input alphabet $\mathcal{X}$ to the
set of density operators in $\mathcal{H}$ (the weak continuity means
continuity of all matrix elements $\langle \psi |\,S_{x}\,|\phi \rangle
;\psi ,\phi \in \mathcal{H}$). Note that according to \cite{del} weak
convergence in the set of density operators is
equivalent to the trace norm convergence. For arbitrary Borel measure $\pi $
on $\mathcal{X}$ we define \begin{equation} \bar{S}_{\pi
}=\int_{\mathcal{X}}S_{x}\pi (dx).  \label{2-5} \end{equation} Because of the
continuity of the function $S_{x}$ the integral is well defined and
represents a density operator in $\mathcal{H}$. Assuming that $H(
\bar{S}_{\pi })<\infty $, we have
\begin{equation}
H(\bar{S}_{\pi })-\int_{\mathcal{X}}H(S_{x})\pi (dx)=\int_{\mathcal{X}
}H(S_{x};\bar{S}_{\pi })\pi (dx)\geq 0,  \label{2-4}
\end{equation}
where the functions $H(S_{x}),H(S_{x};\bar{S}_{\pi })$ ($H(\cdot ;\cdot )$
denotes the quantum relative entropy) are nonnegative and lower
semicontinuous \cite{wehrl}, and hence the integrals are well defined. We
assume that the function $f$ \ is Borel and consider the set $\mathcal{P}
^{B} $ of Borel probability measures $\pi $ on $\mathcal{X}$ satisfying
\begin{equation}
\int_{\mathcal{X}}f(x)\pi (dx)\leq E.  \label{2-0}
\end{equation}

\textbf{Proposition 2.} Let the function $f$ be lower semicontinuous and
tend to infinity at infinity and let there exist a selfadjoint operator $F$
satisfying
\begin{equation}
\mathrm{Tr}\exp \left( -\beta F\right) <\infty \qquad \mathrm{for~all}\quad
\beta >0,  \label{beta}
\end{equation}
such that
\begin{equation}
f(x)\geq \mathrm{Tr}S_{x}F,\quad x\in \mathcal{X}.  \label{fge}
\end{equation}
Then $C$ is finite and
\begin{equation}
C=\max_{\pi \in \mathcal{P}^{B}}\left[ H(\bar{S}_{\pi })-\int_{\mathcal{X}
}H(S_{x})\pi (dx)\right] .
\end{equation}

\textbf{Proof.} The condition (\ref{beta}) implies that the spectrum of $F$
is bounded from below; for simplicity we assume that $F\geq 0$ but the
general case can be reduced to that one. Then the right hand side of 
(\ref{fge}) is defined as in (\ref{TrSF}) below. Denoting
$$S_{\beta }=\left[
\mathrm{Tr}\exp \left( -\beta F\right) \right] ^{-1}\exp \left( -\beta
F\right),$$ we have
\begin{equation}
\beta \mathrm{Tr}\bar{S}_{\pi }F-H(\bar{S}_{\pi })=H(\bar{S}_{\pi };S_{\beta
})-\log \mathrm{Tr}\exp \left( -\beta F\right) ,  \label{frepi}
\end{equation}
whence, by using (\ref{fge}),
\begin{equation*}
H(\bar{S}_{\pi })\leq \beta \mathrm{Tr}\bar{S}_{\pi }F+\log \mathrm{Tr}\exp
\left( -\beta F\right) \leq \beta E+\log \mathrm{Tr}\exp \left( -\beta
F\right) ,
\end{equation*}
hence the condition (\ref{1-62}) is fulfilled, $C$ is finite and equal to 
(\ref{1a-9}). Under the assumptions that the mapping $x\rightarrow S_{x}$ is
weakly continuous, the function $f$ is lower semicontinuous and the
condition (\ref{1-62}) holds, it follows from Proposition 2 of \cite{hol}
that
\begin{equation}
C=\sup_{\pi \in \mathcal{P}^{B}}\left[ H(\bar{S}_{\pi })-\int_{\mathcal{X}
}H(S_{x})\pi (dx)\right] ,  \label{2-6}
\end{equation}
and we wish to prove that the supremum is attained.

In the set of all Borel probability measures on $\mathcal{X}$ we consider
the topology of weak convergence: the sequence $\pi ^{(l)}(dx)$ weakly
converges to $\pi (dx)$ if
\begin{equation*}
\int_{\mathcal{X}}g(x)\pi ^{(l)}(dx)\rightarrow \int_{\mathcal{X}}g(x)\pi
(dx)
\end{equation*}
for all bounded continuous functions $g$ on $\mathcal{X}$. Then one can show
that the set $\mathcal{P}^{B}$ is compact, by using the general criterion
\cite{alex}: a subset $\mathcal{P}^{\prime }$ of  Borel probability measures
on $\mathcal{X}$ is weakly relatively compact iff for any $\varepsilon >0$
there is a compact $\mathcal{K\subset X}$ such that $\pi (\mathcal{X}
\setminus \mathcal{K})\leq \varepsilon $ for all $\pi \in \mathcal{P}
^{\prime }.$ Then $\pi (\mathcal{X}\setminus \mathcal{K})\leq E/\inf_{x\in
\mathcal{X}\setminus \mathcal{K}}f(x)$ for $\pi \in \mathcal{P}^{B},$ which
can be made arbitrarily small.

The map $\pi \rightarrow \bar{S}_{\pi }$ is continuous in the weak operator
topology, and hence, in the trace norm topology. By using this fact we can
prove that the function in the squared brackets of (\ref{2-6}) is upper
semicontinuous and hence attains its maximum on $\mathcal{P}^{B}$. Consider
the first term in the formula (\ref{2-4}). The quantum entropy is lower
semicontinuous, hence the function $\pi \rightarrow H(\bar{S}_{\pi })$ is
lower semicontinuous. Let us show that it is upper semicontinuous and hence
continuous on the set $\mathcal{P}^{B}$. By (\ref{frepi}), (\ref{fge}) we
have
\begin{equation*}
H(\bar{S}_{\pi })\geq \lim \sup_{n\rightarrow \infty }H(\bar{S}_{\pi
^{n}})-\beta \lim \sup_{n\rightarrow \infty }\mathrm{Tr}\bar{S}_{\pi
^{n}}F\geq \lim \sup_{n\rightarrow \infty }H(\bar{S}_{\pi ^{n}})-\beta E,
\end{equation*}
for arbitrary sequence $\{\pi ^{n}\}\in \mathcal{P}^{B}$ weakly converging
to $\pi $. Letting $\beta \rightarrow 0,$ we get the upper semicontinuity.

The second term in (\ref{2-4}) is upper semicontinuous as the greatest lower
bound of continuous functions $\pi \rightarrow -\int g(x)\pi (dx)$, where $g$
varies over bounded continuous functions satisfying $0\leq g(x)\leq
H(S_{x}),\quad x\in \mathcal{X}$. Hence (\ref{2-4}) is upper semicontinuous
and the statement follows. QED

\bigskip \textbf{2.} Now let $\Phi $ be a (quantum-quantum) channel in a
Hilbert space $\mathcal{H}$, i. e. a trace-preserving completely positive
map on trace-class operators in $\mathcal{H}$. We wish to define the
capacity of this channel under additive constraint at the input of the
channel. Let $F$ be positive selfadjoint nonconstant (i. e. not a multiple
of the identity), in general unbounded operator in $\mathcal{H}$,
representing observable the mean value of which is to be constrained (e. g.
energy of the system). For arbitrary density operator $S$ with the spectral
decomposition $S=\sum_{j=1}^{\infty }\lambda _{j}|e_{j}\rangle \langle
e_{j}| $ we define
\begin{equation}
\mathrm{Tr}SF:=\sum_{j=1}^{\infty }\lambda _{j}||\sqrt{F}e_{j}||^{2}\leq
+\infty ,  \label{TrSF}
\end{equation}
assuming $||\sqrt{F}e_{j}||=+\infty $ if $e_{j}$ is not in the domain of  $
\sqrt{F}.$ We impose the analog of the condition (\ref{1-62}):
\begin{equation}
\sup_{S:\mathrm{Tr}SF\leq E}H(\Phi \lbrack S])<\infty ,  \label{uslovie}
\end{equation}
where $E$ is a positive constant.

For the channel $\Phi ^{\otimes n}$ in $\mathcal{H}^{\otimes n}$ the
corresponding observable is
\begin{equation*}
F^{(n)}=F\otimes \dots \otimes I+\dots +I\otimes \dots \otimes F.
\end{equation*}
We want the input states $S^{(n)}$ of the channel $\Phi ^{\otimes n}$
satisfy the additive constraint
\begin{equation}
\mathrm{Tr}S^{(n)}F^{(n)}\leq nE.  \label{constraint}
\end{equation}
Note that (\ref{uslovie}) implies similar property of the channel $\Phi
^{\otimes n}:$
\begin{equation}
\sup_{S^{(n)}:\mathrm{Tr}S^{(n)}F^{(n)}\leq nE}H(\Phi ^{\otimes
n}[S^{(n)}])<\infty .  \label{nuslovie}
\end{equation}
Indeed, by subadditivity of quantum entropy with respect to tensor products,
\begin{equation*}
H(\Phi ^{\otimes n}[S^{(n)}])\leq \sum_{k=1}^{n}H(\Phi \lbrack S_{k}^{(n)}]),
\end{equation*}
where $S_{k}^{(n)}$ is the $k$-th partial state of $S^{(n)}$. Also by
concavity of the entropy
\begin{equation*}
\sum_{k=1}^{n}H(\Phi \lbrack S_{k}^{(n)}])\leq nH(\Phi \lbrack \bar{S}
^{(n)}]),
\end{equation*}
where $\ \ \bar{S}^{(n)}=\frac{1}{n}\sum_{k=1}^{n}S_{k}^{(n)}$. The
inequality (\ref{constraint}) can be rewritten as
\begin{equation*}
\frac{1}{n}\sum_{k=1}^{n}\mathrm{Tr}S_{k}^{(n)}F=\mathrm{Tr}\bar{S}
^{(n)}F\leq E,
\end{equation*}
which implies that
\begin{equation*}
\sup_{S^{(n)}:\mathrm{Tr}S^{(n)}F^{(n)}\leq nE}H(\Phi ^{\otimes
n}[S^{(n)}])\leq n\sup_{S:\mathrm{Tr}SF\leq E}H(\Phi \lbrack S]).
\end{equation*}

\textbf{Definition}. We call by \textit{code} $(\Sigma ^{(n)},M^{(n)})$ of
length $n$ and of size $N$ the collection $\Sigma
^{(n)}=\{S_{w}^{(n)};w=1,\dots ,N\}$ of states satisfying (\ref{constraint}
), with an observable $M^{(n)}=\{M_{j}^{(n)};j=0,1,\dots ,N\}$ in 
$\mathcal{\ H }^{{\otimes n}}$. The \textit{error probability} for the code is
\begin{equation*}
P_{e}(\Sigma ^{(n)},M^{(n)})=\max_{w=1,\dots ,N}\left\{ 1-\mathrm{Tr}\Phi
^{\otimes n}[S_{w}^{(n)}]M_{w}^{(n)}\right\} ,
\end{equation*}
and the minimal error probability over all codes of the length $n$ and the
size $N$ is denoted $p_{e}(n,N)$. The \textit{classical capacity} $C(\Phi )$
is the least upper bound of the rates $R$ for which $\liminf_{n\rightarrow
\infty}p_{e}(n,2^{nR})= 0$.

Let us denote by $\frak{S}^{(n)}$ the set of states in $\mathcal{H}^{\otimes
n}$ satisfying (\ref{constraint}), and by $\mathcal{P}^{(n)}$ the collection
of couples $\left( \pi ^{(n)},\Sigma ^{(n)}\right) ,$ where $\pi _{w}^{(n)}$
are probabilities for the states $S_{w}^{(n)},$ satisfying
\begin{equation}
\sum_{w=1}^{N}\pi _{w}^{(n)}\mathrm{Tr}S_{w}^{(n)}F^{(n)}\leq nE.
\label{constraint2}
\end{equation}

If a probability distribution $\pi ^{(n)}=\{\pi _{w}^{(n)}\}$ on the input
codewords $S_{w}^{(n)}$ is given, then using the transition probability $
p(j|w)=\mathrm{Tr}\Phi ^{\otimes n}[S_{w}^{(n)}]M_{j}^{(n)}$ we can find the
joint distribution of input and output, compute the Shannon information $
\mathcal{I}_{n}(\pi ^{(n)},\Sigma ^{(n)},M^{(n)})$, and define the quantity
\begin{equation*}
\bar{C}^{(n)}(\Phi )=\sup_{\left( \pi ^{(n)},\Sigma ^{(n)}\right) \in
\mathcal{\ P}^{(n)}}\left[ H\left( \sum_{w}\pi _{w}^{(n)}\Phi ^{\otimes
n}[S_{w}^{(n)}]\right) -\sum_{w}\pi _{w}^{(n)}H\left( \Phi ^{\otimes
n}[S_{w}^{(n)}]\right) \right] .
\end{equation*}
If $\Sigma ^{(n)}\subset \frak{S}^{(n)}$, then $\left( \pi ^{(n)},\Sigma
^{(n)}\right) \in \mathcal{\ P}^{(n)}$, and
\begin{equation}
\mathcal{I}_{n}(\pi ^{(n)},\Sigma ^{(n)},M^{(n)})\leq \bar{C}^{(n)}(\Phi ),
\label{qeb}
\end{equation}
by the quantum entropy bound \cite{hol}.

\textbf{Proposition 3}. Let the channel $\Phi $ satisfy the condition 
(\ref{uslovie}). Then the classical capacity of this channel under the constraint
(\ref{constraint}) is finite and equals to
\begin{eqnarray}
C(\Phi ) &=&\lim_{n\rightarrow \infty }\frac{1}{n}\sup_{\pi ^{(n)},\Sigma
^{(n)}\subset \frak{S}^{(n)},M^{(n)}}\mathcal{I}_{n}\left( \pi ^{(n)},\Sigma
^{(n)},M^{(n)}\right)  \label{ineq} \\
&=&\lim_{n\rightarrow \infty }\frac{1}{n}\bar{C}^{(n)}(\Phi ).  \label{eq}
\end{eqnarray}

\textbf{Proof.} Relation (\ref{ineq}) follows from the classical coding
theorem. Inequality $\leq $ in (\ref{eq}) follows then from (\ref{qeb}). Let
us show that
\begin{equation}
C(\Phi )\geq \lim_{n\rightarrow \infty }\frac{1}{n}\bar{C}^{(n)}(\Phi
)\equiv \bar{C}(\Phi ).  \label{concap}
\end{equation}

Take $R<\bar{C}(\Phi )$, then we can choose $n_{0}$, probability
distribution $\pi ^{(n_{0})}=\{\pi _{w}^{(n_{0})}\}$ and collection of
states $\Sigma ^{(n_{0})}=\{S_{w}^{(n_{0})}\}$ in $\mathcal{H}^{\otimes
n_{0}}$ such that $(\pi ^{(n_{0})},\Sigma ^{(n_{0})})\in \mathcal{P}
^{(n_{0})}$ and
\begin{equation}
n_{0}R<H\left( \sum_{w}\pi _{w}^{(n_{0})}\Phi ^{\otimes
n_{0}}[S_{w}^{(n_{0})}]\right) -\sum_{w}\pi _{w}^{(n_{0})}H\left( \Phi
^{\otimes n_{0}}[S_{w}^{(n_{0})}]\right) .  \label{nrh}
\end{equation}
Consider the c-q channel $\tilde{\Phi}$ in $\mathcal{H}^{\otimes n_{0}}$
given by the formula
\begin{equation*}
{\tilde{\Phi}}[S]=\sum_{w}\Phi ^{\otimes n_{0}}[S_{w}]\,\langle
e_{w}|Se_{w}\rangle ,
\end{equation*}
and define the constraint function for this channel as $f(w)=\mathrm{Tr}
S_{w}^{(n_{0})}F^{(n_{0})}.$ The condition (\ref{nuslovie}) implies
\begin{equation*}
\sup_{\pi }H\left( \sum_{w}\pi _{w}\Phi ^{\otimes n_{0}}[S_{w}]\right)
<\infty ,
\end{equation*}
where the supremum is over the probability distributions $\pi $, satisfying
\begin{equation}  \label{rest}
\sum_{w}\pi _{w}f(w)\leq n_{0}E.
\end{equation}
that is, the condition (\ref{1-62}). By the Proposition 1, the capacity of $
\tilde{\Phi}$ is
\begin{equation*}
{C}(\tilde{\Phi})=\sup_{\pi }\left\{ H\left( \sum_{w}\pi _{w}\Phi ^{\otimes
n_{0}}[S_{w}^{(n_{0})}]\right) -\sum_{w}\pi _{w}H(\Phi ^{\otimes
n_{0}}[S_{w}^{(n_{0})}])\right\} ,
\end{equation*}
where the states are fixed and the supremum is over the probability
distributions $\pi $, satisfying (\ref{rest}). By (\ref{nrh}) this is
greater than $n_{0}R$. Denoting $\tilde{p}_{e}(n,N)$ the minimal error
probability for $\tilde{\Phi}$, we have
\begin{equation}
p_{e}(nn_{0},2^{(nn_{0})R})\leq \tilde{p}_{e}(n,2^{n(n_{0}R)}),
\label{errors}
\end{equation}
since every code of size $N$ for $\tilde{\Phi}$ is also code of the same
size for $\Phi $. Indeed, if $\tilde{w}=(w_{1},\dots ,w_{n})$ is a codeword
for $\tilde{\Phi},$ it satisfies the constraint $f(w_{1})+\dots
+f(w_{n})\leq nn_{0}E.$ Defining the state $S_{\tilde{w}
}^{(nn_{0})}=S_{w_{1}}^{(n_{0})}\otimes \dots \otimes S_{w_{n}}^{(n_{0})},$
we see that this is equivalent to $\mathrm{Tr}S_{\tilde{w}
}^{(nn_{0})}F^{(nn_{0})}\leq nn_{0}E,$ that is to the constraint 
(\ref{constraint}) for the q-q channel $\Phi ^{\otimes nn_{0}}.$ Thus having
chosen $R<\bar{C}(\Phi ),$ we can make the right and hence the left hand
side of (\ref{errors}) tend to zero as $n\rightarrow \infty .$ This proves 
(\ref{concap}). QED

These estimates rise questions, to which there is no answer at present. One
may ask whether the additivity $\bar{C}^{(n)}(\Phi )=n\bar{C}^{(1)}(\Phi )$
holds, in which case $C(\Phi )=\bar{C}^{(1)}(\Phi )$. This question looks
even harder than the still unsettled additivity problem in the case of
unconstrained inputs (see \cite{wint} for comments on this problem). The
quantity
\begin{equation}
\bar{C}^{(1)}(\Phi )=\sup_{\sum_{i}\pi _{i}\mathrm{Tr}S_{i}F\leq E}\left[
H\left( \sum_{i}\pi _{i}\Phi \lbrack S_{i}]\right) -\sum_{i}\pi _{i}H\left(
\Phi \lbrack S_{i}]\right) \right]  \label{oneshotunassist}
\end{equation}
looks tractable, although even for the simplest quantum Gaussian channel 
(\ref{gaus}) there is only a natural conjecture about its value and the
solution of the maximization problem (see Subsection 12.6.1 of \cite{hw}).
\bigskip

\textbf{3.} Let us now turn to the entanglement-assisted capacity.
Consider the following protocol of the classical information transmission
through the channel $\Phi $. Systems $A$ and $B$ share an entangled (pure)
state $S_{AB}$. We assume that the amount of entanglement is unlimited but
finite i. e. $H(S_{A})=H(S_{B})<\infty $. $A$ does some encoding $
i\rightarrow \mathcal{E}_{i}$ depending on a classical signal $i$ with
probabilities $\pi _{i}$ and sends its part of this shared state through the
channel $\Phi $ to $B$. Thus $B$ gets the states $(\Phi \otimes \mathrm{Id}
_{B})\left[ S_{i}\right] ,$ where $S_{i}=(\mathcal{E}_{i}\otimes \mathrm{Id}
_{B})\left[ S_{AB}\right] $ with probabilities $\pi _{i}$ and $B$ is trying
to extract the maximum classical information by doing measurements on these
states. Now to enable block coding, all this picture should be applied to
the channel $\Phi ^{\otimes n}.$ Then the signal states $S_{w}^{(n)}$
transmitted through the channel $\Phi ^{\otimes n}\otimes \mathrm{Id}
_{B}^{\otimes n}$ have the special form
\begin{equation}
S_{w}^{(n)}=(\mathcal{E}_{w}^{(n)}\otimes \mathrm{Id}_{B}^{\otimes n})\left[
S_{AB}^{(n)}\right] ,  \label{ABw}
\end{equation}
where $S_{AB}^{(n)}$ is the pure entangled state for $n$ copies of the
system $AB$, satisfying the condition $H(S_{B}^{(n)})<\infty $, and $
w\rightarrow \mathcal{E}_{w}^{(n)}$ are the encodings for $n$ copies of the
system $A.$ We impose the constraint (\ref{constraint}) onto the input
states of the channel $\Phi ^{\otimes n}$, which is equivalent to similar
constraint for the channel $\Phi ^{\otimes n}\otimes \mathrm{Id}
_{B}^{\otimes n}$ with the constraint operators $F_{AB}^{(n)}=F^{(n)}\otimes
I_{B}^{\otimes n}.$ We denote by $\mathcal{P}_{AB}^{(n)}$ the collection of
couples $(\pi ^{(n)},\Sigma ^{(n)}),$ where $\pi ^{(n)}=\{\pi _{w}^{(n)}\}$
is the probability distribution and $\Sigma ^{(n)}=\{S_{w}^{(n)}\}$ is the
collection of states of the form (\ref{ABw}) satisfying the constraint 
(\ref{constraint2}) with the operators $F_{AB}^{(n)}.$ The classical capacity of
this protocol will be called \textit{entanglement-assisted classical
capacity $C_{ea}(\Phi )$ of the channel $\Phi $ under the constraint} 
(\ref{constraint}).

Let $S$ be a density operator such that both $H(S)$ and $H(\Phi (S))$ are
finite, then the \textit{quantum mutual information} is
\begin{equation}
I(S,\Phi )=H(S)+H(\Phi (S))-H(S;\Phi )  \label{qmi}
\end{equation}
where $H(S;\Phi )$ is the entropy exchange (see e. g. \cite{hw}). If the
constraint operator $F$ satisfies (\ref{beta}), then $H(S)$ is finite for
all $S$ satisfying $\mathrm{Tr}SF\leq E$. Indeed, we have
\begin{equation}
\beta \mathrm{Tr}SF-H(S)=H(S;S_{\beta })-\log \mathrm{Tr}\exp \left( -\beta
F\right) ,  \label{free}
\end{equation}
hence
\begin{equation}
H(S)\leq \beta E+\log \mathrm{Tr}\exp \left( -\beta F\right).  \label{us}
\end{equation}

\textbf{Proposition 4}. Let $\Phi $ be a channel satisfying the condition 
(\ref{uslovie}) with the operator $F$ satisfying (\ref{beta}), then its
entanglement-assisted classical capacity under the constraint (\ref
{constraint}) is finite and equals to
\begin{equation}
C_{ea}(\Phi )=\sup_{S:\mathrm{Tr}SF\leq E}I\left( S,\Phi \right) .
\label{eaco}
\end{equation}

\textbf{Proof}. By a modification of the proof of Proposition 2, we have
\begin{equation}
C_{ea}(\Phi )=\lim_{n\rightarrow \infty }\frac{1}{n}C_{ea}^{(n)}(\Phi ),
\label{cea}
\end{equation}
where
\begin{equation*}
C_{ea}^{(n)}(\Phi )=\sup_{\left( \pi ^{(n)},\Sigma ^{(n)}\right) \in
\mathcal{\ }\mathcal{P}_{AB}^{(n)}}\biggl[ H\left( \sum_{w=1}^{N}\pi
_{w}^{(n)}\left( \Phi ^{\otimes n}\otimes \mathrm{Id}_{B}^{\otimes n}\right)
[S_{w}^{(n)}]\right)
\end{equation*}
\begin{equation}
-\sum_{w=1}^{N}\pi _{w}^{(n)}H\left( \left( \Phi ^{\otimes n}\otimes \mathrm{
\ Id}_{B}^{\otimes n}\right) [S_{w}^{(n)}]\right) \biggr] .  \label{cean}
\end{equation}
Note that all terms in squared brackets are finite because of the assumed
finiteness of the entropy $H(S^{(n)}_B)$ and (\ref{uslovie}).

We first prove the inequality $\leq $ in (\ref{eaco}). By using (\ref{cea}),
(\ref{cean}) and the inequality (17) from \cite{heac} we obtain
\begin{equation*}
C_{ea}(\Phi )\leq \lim_{n\rightarrow \infty }\frac{1}{n}\sup_{(\pi
^{(n)},\Sigma ^{(n)})\in \mathcal{P}_{AB}^{(n)}}I\left( \sum_{w=1}^{N}\pi
_{w}^{(n)}\mathrm{Tr}_{B^n}S_{w}^{(n)},\Phi ^{\otimes n}\right) .
\end{equation*}
The right hand side is less than or equal to
\begin{equation*}
\sup_{S^{(n)}:\mathrm{Tr}S^{(n)}F^{(n)}\leq nE}I\left( S^{(n)},\Phi
^{\otimes n}\right) \equiv \bar{I}_{n}(\Phi ).
\end{equation*}
But the sequence $\bar{I}_{n}(\Phi )$ is additive; it is sufficient to prove
only
\begin{equation}
\bar{I}_{n}(\Phi )\leq n\bar{I}_{1}(\Phi ).  \label{subad}
\end{equation}
Indeed, by subadditivity of quantum mutual information,
\begin{equation*}
I\left( S^{(n)},\Phi ^{\otimes n}\right) \leq \sum_{j=1}^{n}I\left(
S_{j}^{(n)},\Phi \right) ,
\end{equation*}
where $S_{j}^{(n)}$ are the partial states, and by concavity,
\begin{equation*}
\sum_{j=1}^{n}I\left( S_{j}^{(n)},\Phi \right) \leq n\sum_{j=1}^{n}I\left(
\frac{1}{n}\sum_{j=1}^{n}S_{j}^{(n)},\Phi \right) .
\end{equation*}
But $\mathrm{Tr} S^{(n)}F^{(n)}\leq nE$ is equivalent to $\mathrm{Tr}\left(
\frac{1}{n} \sum_{j=1}^{n}S_{j}^{(n)}\right) F\leq E,$ hence (\ref{subad})
follows. Thus
\begin{equation}
C_{ea}(\Phi )\leq \sup_{S:\mathrm{Tr}SF\leq E}I\left( S,\Phi \right) .
\label{estmui}
\end{equation}

The proof of the converse inequality is based on the expression (\ref{cean})
and the specific encoding protocol from \cite{eac2}, \cite{heac}.

Since $F$ is nonconstant operator, the image of the convex set of all
density operators under the map $S\rightarrow \mathrm{Tr}SF$ is an interval.
Assume first that $E$ is not the minimal eigenvalue of \ $F.$ Then there
exist a real number $E^{\prime }$ and a density operator $S$ in $\mathcal{H}
_{A}$ such that $\mathrm{Tr}SF\leq E^{\prime }<E.$ Let $S=\sum_{j=1}^{\infty
}\lambda _{j}|e_{j}\rangle \langle e_{j}|$ be its spectral decomposition,
and define $S_{d}=\sum_{j=1}^{d}\tilde{\lambda}_{j}|e_{j}\rangle \langle
e_{j}|,$ where $\tilde{\lambda}_{j}=\left( \sum_{k=1}^{d}\lambda _{k}\right)
^{-1}\lambda _{j}.$ Then $\left\| S-S_{d}\right\| _{1}\rightarrow 0$ as $
d\rightarrow \infty ,$ where $\left\| \cdot \right\| _{1}$ is the trace
norm. Denote $f(j)=||\sqrt{F}e_{j}||^{2},$ then
\begin{equation*}
\mathrm{Tr}S_{d}F=\sum_{j=1}^{d}\tilde{\lambda}_{j}f(j)=E^{\prime
}+\varepsilon _{d},
\end{equation*}
where $\ \varepsilon _{d}\rightarrow 0$ as $d\rightarrow \infty .$ Now
consider the density operator $S_{d}^{\otimes n}$, denote by $P^{n,\delta }$
its strongly typical projector \cite{heac}\ and let $d_{n,\delta }=\dim
P^{n,\delta },$ $\bar{S}_{d}^{n,\delta }=\frac{P^{n,\delta }}{d_{n,\delta }}.
$ Due to the strong typicality, we have similarly to the estimate at the
bottom of p. 4329 in \cite{heac}
\begin{equation*}
\left| \mathrm{Tr}\left( \bar{S}_{d}^{n,\delta }-S_{d}^{\otimes n}\right)
F^{(n)}\right| \leq n\delta \max \left\{ f(j);j=1,\dots ,d\right\} ,
\end{equation*}
whence
\begin{equation*}
\mathrm{Tr}\bar{S}_{d}^{n,\delta }F^{(n)}\leq \mathrm{Tr}S_{d}^{\otimes
n}F^{(n)}+n\delta \max \left\{ f(j);j=1,\dots ,d\right\}
\end{equation*}
\begin{equation*}
=n\left( E^{\prime }+\varepsilon _{d}+\delta \max \left\{ f(j);j=1,\dots
,d\right\} \right) .
\end{equation*}

For every $d$ large enough one can find $\delta _{0}$ such that the right
hand side is $\leq nE$ for $\delta \leq \delta _{0}.$ Then using the
expression (\ref{cean}) and the aforementioned encoding protocol, we can
prove similarly to \cite{eac2} or to (7) in \cite{heac}:
\begin{equation*}
C_{ea}^{(n)}(\Phi )\geq I\left( \bar{S}_{d}^{n,\delta },\Phi ^{\otimes
n}\right) .
\end{equation*}
Indeed, take the classical signal to be transmitted as $w=(\alpha ,\beta );$
$\alpha ,\beta =1,\dots ,d_{n,\delta }$ with equal probabilities $\pi
_{w}=1/d_{n,\delta }^{2},$ the maximally entangled state $S_{AB}=|\psi
_{AB}\rangle \langle \psi _{AB}|$ and the unitary encodings $\mathcal{\ E}
_{A}^{w}\left[ S\right] =W_{\alpha \beta }SW_{\alpha \beta }^{\ast }$ (see
the proof of Theorem in \cite{heac}). Such an encoding satisfies the input
constraint because
\begin{equation*}
\sum_{w}\pi _{w}\mathcal{E}_{A}^{w}\left[ S_{AB}\right] =\bar{S}
_{d}^{n,\delta }\otimes \bar{S}_{d}^{n,\delta }.
\end{equation*}
Thus for this protocol the condition $\left( \pi ^{(n)},\Sigma
^{(n)}\right) \in \mathcal{P}_{AB}^{(n)}$ in (\ref{cean}) is equivalent to $
\mathrm{Tr}\bar{S}_{d}^{n,\delta }F^{(n)}\leq nE.$

Passing to the limit $n\rightarrow \infty ,\delta \rightarrow 0,$ and using
the approximation argument from \cite{heac} we obtain
\begin{equation*}
C_{ea}(\Phi )=\lim_{n\rightarrow \infty }\frac{1}{n}C_{ea}^{(n)}(\Phi )\geq
I\left( S_{d},\Phi \right) ,
\end{equation*}
where $\mathrm{Tr}S_{d}F=E^{\prime }+\varepsilon _{d}\leq E.$ Finally, we
pass to the limit $d\rightarrow \infty $ and show that
\begin{equation}
\liminf_{d\rightarrow \infty }I\left( S_{d},\Phi \right) \geq I\left( S,\Phi
\right) .  \label{lower}
\end{equation}
To see it, we represent the mutual information as quantum relative entropy
\begin{equation}
I\left( S,\Phi \right) =H\left( \left( \Phi \otimes \mathrm{Id}_{R}\right)
\left[ |\psi \rangle \langle \psi |\right] ;\Phi \left[ S\right] \otimes
S\right) ,  \label{e}
\end{equation}
where $|\psi \rangle \langle \psi |$ is a purification for $S$, $R$ is a
purifying system, and similarly for $I(S_{d},\Phi ).$ If $|\psi \rangle
=\sum_{j=1}^{\infty }\sqrt{\lambda _{j}}|e_{j}\rangle \otimes |e_{j}\rangle
, $ then
\begin{equation*}
|\psi _{d}\rangle =\sum_{j=1}^{d}\sqrt{\tilde{\lambda}_{j}}|e_{j}\rangle
\otimes |e_{j}\rangle
\end{equation*}
is a purification for $S_{d}.$ We have $\left\| |\psi \rangle -|\psi
_{d}\rangle \right\| \rightarrow 0,$ and hence
\begin{equation*}
\left\| |\psi \rangle \langle \psi |-|\psi _{d}\rangle \langle \psi
_{d}|\right\| _{1}\rightarrow 0\quad \mathrm{as}\quad d\rightarrow \infty ,
\end{equation*}
therefore (\ref{lower}) follows from the lower semicontinuity of the
relative entropy \cite{wehrl}. Thus we obtain
\begin{equation*}
C_{ea}(\Phi )\geq I\left( S,\Phi \right) ,
\end{equation*}
where $S$ is an arbitrary density operator with $\mathrm{Tr}SF<E.$ This is
easily extended to operators with $\mathrm{Tr}SF=E$ by approximating them
with the operators $S_{\epsilon }=(1-\epsilon )S+\epsilon |e\rangle \langle
e|,$ where $e$ is chosen such that $\langle e|F|e\rangle <E.$

In case $E$ is the minimal eigenvalue of $F,$ the condition $\mathrm{Tr}
SF\leq E$ amounts to the fact that the support of $S$ is contained in the
spectral projection of $F$ corresponding to this minimal eigenvalue. The
condition (\ref{beta}) implies that the eigenvalues of $F$ have finite
multiplicity. Thus the support of $S$ is fixed finite dimensional subspace
and we can take $S_{d}=S.$ Then we can repeat the above argument with the
equality $\mathrm{Tr}SF=E$ holding at each step. To sum up, we have
established

\begin{equation*}
C_{ea}(\Phi )\geq \sup_{S:\mathrm{Tr}SF\leq E}I\left( S,\Phi \right) ,
\end{equation*}
and thus the equality in (\ref{eaco}). QED

\bigskip Now we investigate the question when the supremum in the right hand
side of (\ref{eaco}) is achieved.

\textbf{Lemma}. Let the spectrum of operator $F$ consist of eigenvalues $
f_{n}$ of finite multiplicity and $\lim_{n\rightarrow \infty }
f_{n}=+\infty $, then the set $\frak{S}_{E}:=\left\{ S:\mathrm{Tr}SF\leq
E\right\} $ is compact.

\textbf{Proof}. Without loss of generality we assume that $ f_{n}$ is
monotonously increasing and denote by $P_{n}$ the finite dimensional
projection onto the eigenspace corresponding to the first $n$ eigenvalues,
then $P_{n}\uparrow I.$ By a general criterion, a weakly closed subset $
\frak{S}^{\prime }$ of density operators is weakly compact if and only if
for every $\varepsilon >0$ there is a finite dimensional projection $P$ such
that $\mathrm{Tr}S(I-P)\leq \varepsilon $ for all $S\in \frak{S}^{\prime }$,
see \S III.9 of \cite{sarym}. But according to \cite{del}, the weak
convergence of density operators is equivalent to their trace norm
convergence. Since $f_{n+1}(I-P_{n})\leq \ F,$ we have $\mathrm{Tr}
S(I-P_{n})\leq \ f_{n+1}^{-1}\mathrm{Tr}SF\leq \ f_{n+1}^{-1}E$ for $S\in
\frak{S}_{E},$ whence the lemma follows. QED

Notice that condition (\ref{beta}) implies that $F$ satisfies the condition
of the lemma.

\textbf{Proposition 5.} Let the constraint operator $F$ satisfy the
condition (\ref{beta}), and let there exist a selfadjoint operator $\tilde{F}
$ satisfying (\ref{beta}) such that $\Phi ^{\ast }\left[ \tilde{F}\right]
\leq F,$ where $\Phi ^{\ast }$ is the dual channel. Then
\begin{equation}
C_{ea}(\Phi )=\max_{S:\mathrm{Tr}SF\leq E}I\left( S,\Phi \right) .
\label{aco}
\end{equation}

\textbf{Proof.} We shall treat separately each term in the formula (\ref{qmi}). 
Notice that quantum entropy is lower semicontinuous. Since the entropy
exchange can be represented as $H(S;\Phi )=H(\Phi _{E}\left[ S\right] ),$
where $\Phi _{E}$ is a channel from the system space $\mathcal{H}_{A}$ to
the environment space $\mathcal{H}_{E}$, it is also lower semicontinuous and
thus the last term in (\ref{qmi}) is upper semicontinuous. Concerning the
first term, it is upper semicontinuous and hence continuous on the set $
\frak{S}_{E}$ $=\left\{ S:\mathrm{Tr}SF\leq E\right\} $ if the constraint
operator $F$ satisfies (\ref{beta}). The proof goes as follows: we have
\begin{equation}
\beta \mathrm{Tr}S_{n}F-H(S_{n})=H(S_{n};S_{\beta })-\log \mathrm{Tr}\exp
\left( -\beta F\right) ,  \label{fre}
\end{equation}
and similarly for $S$ instead of $S_{n}.$ By using lower semicontinuity of
the relative entropy, we obtain
\begin{equation*}
H(S)\geq \lim \sup_{n\rightarrow \infty }H(S_{n})-\beta \lim
\sup_{n\rightarrow \infty }\mathrm{Tr}S_{n}F.
\end{equation*}
For $S_{n}\in \frak{S}_{E}$ the last term is $\geq -\beta E,$ which can be
made arbitrarily small.

We can apply similar argument to the second term in (\ref{qmi}) under the
assumption that there exists a selfadjoint operator $\tilde{F}$ satisfying (
\ref{beta}) and such that $\Phi ^{\ast }\left[ \tilde{F}\right] \leq
F; $ the relation (\ref{fre}) is then replaced with
\begin{equation*}
\beta \mathrm{Tr}S_{n}F-H(\Phi \left[ S_{n}\right] )\geq H(\Phi \left[ S_{n}
\right] ;\tilde{S}_{\beta })-\log \mathrm{Tr}\exp \left( -\beta \tilde{F}
\right) ,
\end{equation*}
where $\tilde{S}_{\beta }=\left[ \mathrm{Tr}\exp \left( -\beta \tilde{F}
\right) \right] ^{-1}\exp \left( -\beta \tilde{F}\right) ,$ and the proof
goes in a similar way. Moreover, this assumption implies also that the
condition (\ref{uslovie}) and hence (\ref{eaco}) holds. Indeed, denoting $
\Phi \lbrack S]=S^{\prime }$ and using (\ref{us}) we have
\begin{equation*}
\sup_{S:\mathrm{Tr}SF\leq E}H(\Phi \lbrack S])\leq \sup_{S^{\prime }:\mathrm{
Tr}S^{\prime }\tilde{F}\leq E}H(S^{\prime })\leq \beta E+\log \mathrm{Tr}
\exp \left( -\beta \tilde{F}\right) .
\end{equation*}
QED

This set of conditions ensuring that the supremum in (\ref{eaco}) is
achieved, is fulfilled for example in the case where $\Phi $ is a Bosonic
Gaussian channel and $F$ is positive quadratic polynomial in canonical
variables, e. g. energy operator \cite{hw}. The simplest Gaussian channel
''quantum signal plus classical noise'' is described in the Heisenberg
picture by the equation:
\begin{equation}
a\rightarrow a+\xi ,  \label{gaus}
\end{equation}
where $a$ is the annihilation operator of the mode, and $\xi $ is the
classical complex Gaussian random variable with zero mean and the variance $N
$ (the mean photon number in the noise). The constraint is $\mathrm{Tr}
Sa^{\dagger }a\leq $ $E,$ where $S$ is the density operator of the signal $a.
$ The gain of entanglement assistance $G=C_{ea}\left( \Phi \right)
/C^{(1)}\left( \Phi \right) $ was computed in \cite{eac2},  \cite{hw}. In
particular, when the signal mean photon number $E$ tends to zero while $N>0$,
\begin{equation*}
C^{(1)}\left( \Phi \right) \sim E\log \left( \frac{N+1}{N}\right) ,\quad
\quad C_{ea}\left( \Phi \right) \sim -E\log E/(N+1),
\end{equation*}
and $G$ tends to infinity as $-\log E$.

\bigskip In this paper we were interested in the situation where all the
entropy terms entering the expressions for the capacities are finite, which
was ensured by the conditions (\ref{uslovie}), (\ref{beta}). Taking this as
an approximation, one can obtain in the general case expressions involving
only relative entropy and thus unambiguously defined with values in the
range $[0,+\infty ]$. For unassisted capacities cf. \cite{hol}, \cite{hana};
$C_{ea}(\Phi )$ will be given by (\ref{eaco}), where $I\left( S,\Phi \right)
$ is defined as in (\ref{e}).

\end{document}